\documentclass[prc,twocolumn,showpacs]{revtex4}
\usepackage{graphicx}
\usepackage{float}
\newcommand{\D}{\displaystyle}

\begin{document}
\title {The stability and the shape of the heaviest nuclei}

\author{
L. S. Geng,$^{1,2,3}$ H. Toki$^{2}$ and E. G. Zhao$^3$}
\affiliation{
$^1$School of Physics, Peking University, Beijing 100871, People's
Republic of China \\
$^2$Research Center for Nuclear Physics (RCNP), Osaka University,
Ibaraki 567-0047, Japan\\
$^{3}$Institute of Theoretical Physics, Chinese Academy of
Sciences, Beijing 100080, China}
\date{\today}

\begin{abstract}
In this paper, we report a systematic study of the heaviest nuclei
within the relativistic mean field (RMF) model. By comparing our
results with those of the Hartree-Fock-Bogoliubov method (HFB) and
the finite range droplet model (FRDM), the stability and the shape
of the heaviest nuclei are discussed. The theoretical predictions
as well as the existing experimental data indicate that the
experimentally synthesized superheavy nuclei are in between the
fission stability line, the line connecting the nucleus with
maximum binding energy per nucleon in each isotopic chain, and the
$\beta$-stability line, the line connecting the nucleus with
maximum binding energy per nucleon in each isobaric chain. It is
shown that both the fission stability line and the
$\beta$-stability line tend to be more proton rich in the
superheavy region. Meanwhile, all the three theoretical models
predict most synthesized superheavy nuclei to be deformed.
\end{abstract}
\maketitle

\section{Introduction}
Recent claims of successful syntheses of superheavy elements 115
and 113 \cite{Oganessian04,Morita04} have aroused new enthusiasm
about studies of superheavy nuclei in the nuclear physics
community (see Refs. \cite{Geng03PRC,Sharma05,Zhang04} and
references therein). For a review of recent experimental progress
of this subject, we refer the reader to Refs.
\cite{Hofmann98,Hofmann00}. Conventional liquid drop models of
finite nuclei forbid the existence of any nuclei with a proton
number larger than 100, i.e. superheavy nuclei, due to the
destructive Coulomb force. However, shell effects are found to be
able to stabilize these nuclei, and therefore explain their very
existences \cite{Strutinsky68,Ring80}. It has long been predicted
that there exist a large number of relatively long-lived
superheavy nuclei, the so-called superheavy island, which is
separated in neutron and proton numbers from the known heavy
elements by a region of much higher instability. Although the
experimentally-synthesized superheavy nuclei are indeed very
heavy, it is generally believed that they are not examples of the
originally sought island of superheavy elements.

On the theoretical side, a lot of efforts have been made to
interpret the experimental results and make various predictions. A
short review of the theoretical activities can be found in Ref.
\cite{Zhang04}. Nowadays, there are several categories of
theoretical models often used to study superheavy nuclei: The
first category is the liquid drop model and its many variants,
such as the finite-range droplet model (FRDM) \cite{Moller95}; The
second category is the non-relativistic Skyme-Hartree-Fock model;
The third category is the relativistic mean field model. Using
these models, fission barriers, alpha-decay energies, shell
closures, single-particle spectra, and so on have been extensively
discussed \cite{Zhang04}.

In recent years, the relativistic mean field model has received
much attention due to its natural description of the spin-orbit
interaction, the saturation properties of symmetric nuclear
matter, and many other things that non-relativistic models have
some difficulties to explain \cite{Meng05}. It has also been
widely employed to study superheavy nuclei (a short review can be
found in Ref.\cite{Zhang04}). Due to the amount of computer
resources needed, studies of superheavy nuclei in the relativistic
mean field model have often been limited to either spherical
assumption or a small part of the superheavy region.  In this
paper, we report the first systematic study of superheavy nuclei
within the relativistic mean field model with the deformation
effect and the pairing correlation properly treated. By comparing
our RMF+BCS calculations with those of the Hartree-Fock-Bogoliubov
(HFB) and finite range droplet model (FRDM), we hope to obtain
some hints for the search of the superheavy island.

In particular, we would like to adress two interesting subjects:
the stability and the shape of the heaviest nuclei. The stability
of a superheavy nucleus is a very subtle subject. It
 is determined by many competing decay modes: alpha decay, spontaneous fission,
 beta decay, and etc.
 Except beta decay, the other two are very difficult
 to describe quantitively, which would involve complicated lifetime calculations. 
 Beta decay, on the other hand, can be understood much more easily from the energy
 point of view, i.e. nuclei near the $\beta$-stability line are stable against beta decay.
 Similarly, a useful concept is  the fission stability line,
 the line connecting the nucleus with maximum
binding energy per nucleon in each isotopic chain, which is
related to the minimum Q value of fission \cite{Wu96}. At this
line, nuclei with fixed proton number have maximum binding energy
per nucleon; therefore they would be stable against neutron
emission, which can play an important role to synthesize
superheavy nuclei at the first place \cite{Oganessian04,Morita04}.
Hence, one would expect superheavy nuclei should not deviate too
far from this line.

The shape of the heaviest nucleus can also influence its stability
greatly \cite{Bohr98}. The original island of superheavy elements
is predicted to be around $Z=114$ and $N=184$ \cite{Moller94}
mainly due to the fact that $_{184}114$ is predicted to be a
doubly magic nucleus with spherical shape, where shell effect is
the strongest. Recent investigations have provided somehow
conflicting predictions for the next doubly magic system, for
example, the non-relativistic forces SkM* and SkP seem to prefer
$_{184}126$ instead of $_{184}114$ \cite{Rutz97}. A more complete
summary of various predictions of different models can be found in
Ref. \cite{Zhang04}. In this paper, we would like to compare our
RMF+BCS predictions with those of the HFB model and the FRDM model
in order to see whether a large number of spherical nuclei exist
in these models, which would indicate the existence or
nonexistence of the next doubly magic nucleus, or (less
ambitiously) the next neutron (proton) shell closure.

This paper is organized as follows. In Section II, we briefly
introduce the relativistic mean field model and explain the
numerical details of our calculation. In Section III, the
stabilities of superheavy nuclei are studied from the energy point
of view. In Section IV, the shapes of superheavy nuclei predicted
by different theoretical models are compared. The whole work is
summarized in Section V.

\section{Theoretical framework}
In this section, we briefly describe the RMF+BCS calculations. The
RMF+BCS calculations have been carried out using the model
Lagrangian density with nonlinear terms for both $\sigma$ and
$\omega$ mesons as described in detail in Refs. \cite{Geng05PTP,
Sugahara94}, which is given by
\begin{equation}
\begin{array}{lll}
\mathcal{L} &=& \bar \psi (i\gamma^\mu\partial_\mu -M) \psi \\
&+&\,\frac{\D 1}{\D
2}\partial_\mu\sigma\partial^\mu\sigma-\frac{\D 1}{\D
2}m_{\sigma}^{2} \sigma^{2}- \frac{\D 1}{ \D
3}g_{2}\sigma^{3}-\frac{\D 1}{\D
4}g_{3}\sigma^{4}-g_{\sigma}\bar\psi
\sigma \psi\\
&-&\frac{\D 1}{\D 4}\Omega_{\mu\nu}\Omega^{\mu\nu}+\frac{\D 1}{\D
2}m_\omega^2\omega_\mu\omega^\mu +\frac{\D 1}{\D
4}g_4(\omega_\mu\omega^\mu)^2-g_{\omega}\bar\psi
\gamma^\mu \psi\omega_\mu\\
 &-& \frac{\D 1}{\D 4}{R^a}_{\mu\nu}{R^a}^{\mu\nu} +
 \frac{\D 1}{\D 2}m_{\rho}^{2}
 \rho^a_{\mu}\rho^{a\mu}
     -g_{\rho}\bar\psi\gamma_\mu\tau^a \psi\rho^{\mu a} \\
      &-& \frac{\D 1}{\D 4}F_{\mu\nu}F^{\mu\nu} -e \bar\psi
      \gamma_\mu\frac{\D 1-\tau_3}{\D 2}A^\mu
      \psi,\\
\end{array}
\end{equation}
where all symbols have their usual meanings. The corresponding
Dirac equation for nucleons and Klein-Gordon equations for mesons
obtained with the mean-field approximation and the no-sea
approximation are solved by the expansion method on the axially
deformed harmonic-oscillator basis \cite{Gambhir90}. The number of
shells used for expanding the nucleon and meson wave functions is chosen as $N_f=N_b=20$. More shells
have been tested for convergence considerations. Quadrupole
constrained calculations \cite{Flocard73} have been performed for
all the nuclei considered here in order to obtain their energy
surfaces and determine the corresponding ground-state
deformations.

The pairing correlation plays an important role in studies of
open-shell nuclei. It is also true for superheavy nuclei. In
Ref. \cite{Geng03PRC}, it was shown that the use of a zero-range
$\delta$-force  in the particle-particle channel can bring down
the fission barrier compared to the use of a constant-gap pairing
method. In the present calculation, the pairing correlation is
treated by a state-dependent BCS method \cite{Geng03PTP}.  More
specifically, the pairing force used is of the volume type
 \begin{equation}
 V=V_0\delta(\vec{r}-\vec{r}').
 \end{equation}
 In the past years, whether the pairing correlation in finite nuclei is of a volume
 type or surface type has been discussed a lot ,
 but it seems that more investigations are still needed to reach a definite
 conclusion \cite{Sandulescu05}; therefore, to limit the number of free parameters,
 we do not introduce explicitly any density dependence. On the other hand,
 to describe simultaneously both light
 and heavy nuclei \cite{Geng05PTP}, we introduce a weak mass number dependence to
 the pairing strength, i.e.
 \begin{equation}
 V_0=300+120/A^{1/3},
 \end{equation}
 which is purely phenomenological except for the $A^{1/3}$ dependence \cite{Satula98},  For nuclei with
 an odd-number of
 nucleons, the blocking effect has been treated within the
BCS framework \cite{Geng03PRC,Ring80}. A more detailed description of the
pairing method  can be found in Ref. \cite{Geng03PRC}.

In the mean field channel, the effective force TMA
\cite{Sugahara94} is used. The effective force TMA was first
proposed to describe simultaneously both light and heavy nuclei.
It in fact originated from two other very successful parameter
sets: TM1 and TM2. TM1 aimed to describe the ground-state
properties of heavy nuclei ($A>40$) and TM2 those of light nuclei
($A<40$). On the one hand, TMA inherited TM1's favorable property
of being able to reproduce the essential feature of the equation
of state and the vector and the scalar self-energies of the
relativistic Bruckner-Hartree-Fock theory for nuclear matter
\cite{Brockmann92}. On the other hand, it can also describe light
nuclei very well. Its success comes from a weak mass dependence,
which smoothly interpolates the TM1 and the TM2 parameter sets. We
may interpret this mass dependence as a mean to effectively
express the quantum fluctuations beyond the mean field level
and/or the softness of the nuclear ground states in deformation,
pairing and alpha clustering in light nuclei. Compared to other
successful effective forces, such as NL3, the description of
finite nuclei is of similar quality or slightly better
\cite{Geng05PTP}, but TMA(TM1) yields a much softer equation of
state at high density, which seems to be favored by current
experimental results.

Finally, we would like to mention the strategy we used to confine our study to a
reasonable number of nuclei since calculations of superheavy nuclei cost a lot of time.
The number of nuclei is determined by including all of those
compiled in Ref. \cite{Audi03} and extending the corresponding
proton-rich limit and neutron-rich limit of a certain isotopic
chain \cite{Audi03} by ten more nuclei, respectively (see also Fig.\ref{mass}). Using such a strategy, the number of
superheavy nuclei investigated in the present study is around 600.

\section{The stability of superheavy elements}
\begin{figure}[t]
\centering
\includegraphics[scale=0.4]{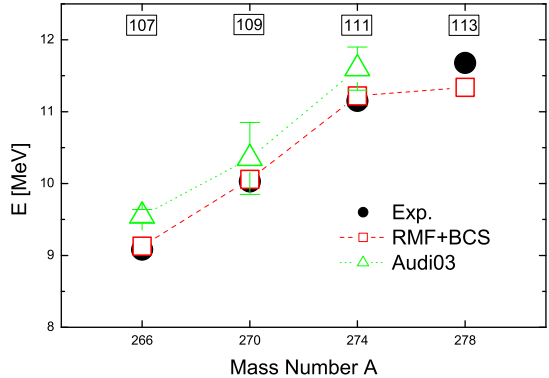}
\caption{(color online) $\alpha$-decay energies of the $^{278}113$
$\alpha$-decay chain. The latest experimental results of Morita et
al. \cite{Morita04} are compared with the predictions of the
RMF+BCS calculation and those compiled in Audi03 (obtained using ``systematic trends")
\cite{Audi03}.
\label{alpha-113}}
\end{figure}

\begin{figure}[htpb]
\centering
\begin{minipage}[c]{0.5\textwidth}
\includegraphics[scale=0.4]{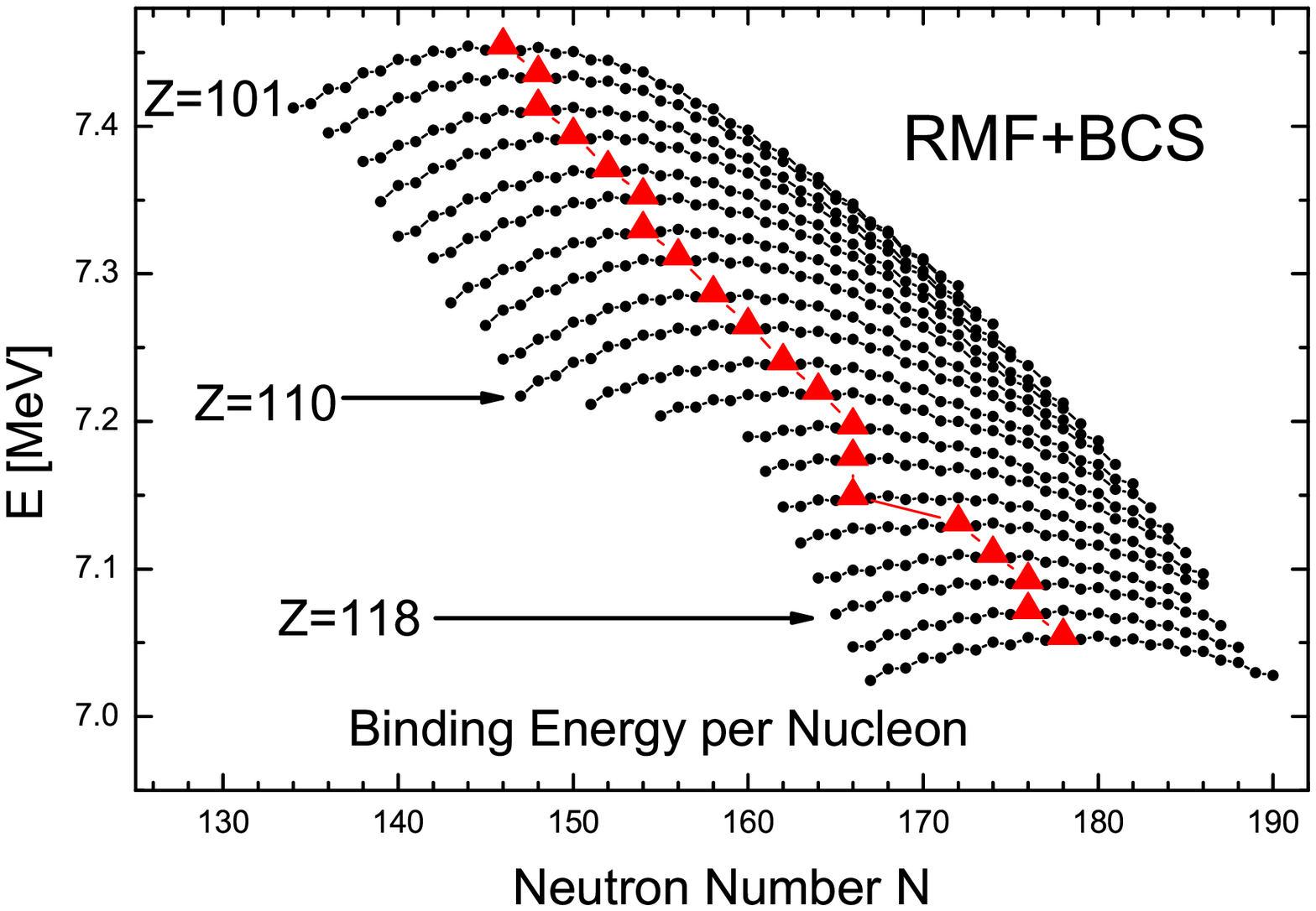}
\end{minipage}
\begin{minipage}[c]{0.5\textwidth}
\includegraphics[scale=0.4]{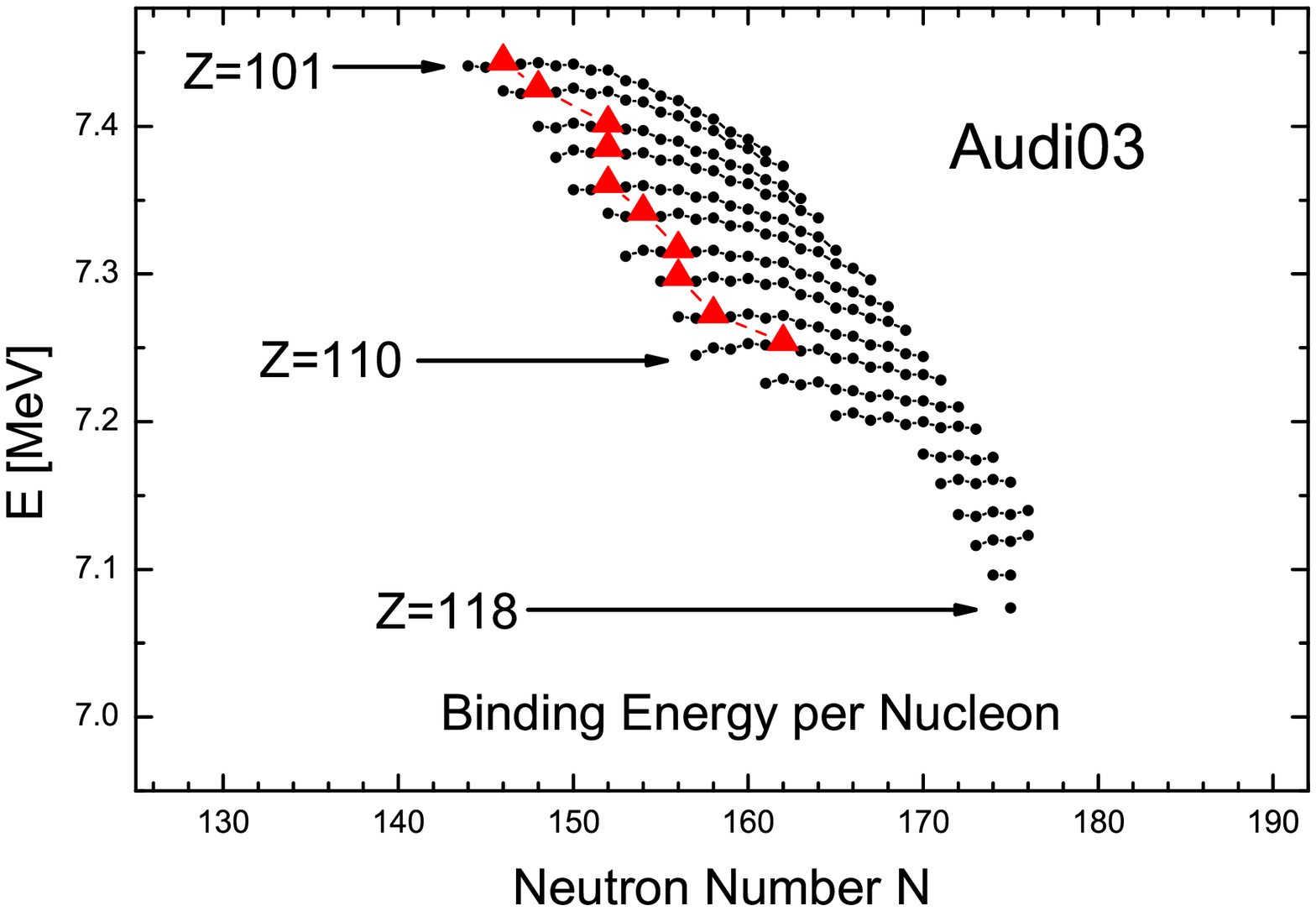}
\end{minipage}
\caption{(color online). Binding energies per nucleon of
superheavy nuclei with $Z = 101$-120 and $N = 134$-190 as
functions of the neutron number $N$. The theoretical predictions
are compared with existing experimental data (including those
obtained using ``systematic trends") \cite{Audi03}. Different
isotopes are connected by solid lines and ordered from top to
bottom with increasing $Z$. The nucleus with maximum binding
energy per nucleon in each isotopic chain is denoted by a
triangle, and those of different isotopic chains are connected by
dashed lines.\label{mass}}
\end{figure}
Before we study the stability of the heaviest nuclei, it is
necessary to stress that although the parameters of the RMF+BCS
model have not been constrained by any information from the
superheavy region, the agreement of its predictions with existing
experimental data is quite good in general, which has been
demonstrated in the entire region (see Refs.
\cite{Meng00PRC,Long02,Ren03,Geng03PRC,Zhang04} and references
therein). Recently, Morita et al. has reported the synthesis of
the new element $113$ \cite{Morita04}. In Fig. \ref{alpha-113},
the experimental $\alpha$-decay energies of the $^{278}$113
$\alpha$-decay chain are compared with our predictions and those
compiled in Audi03 (obtained using ``systematic trends''
\cite{Audi03}. It is clearly seen that the agreement is
remarkable. A more detailed study is underway and will be reported
somewhere else.

\begin{figure*}[htpb]
\centering
\includegraphics[scale=0.6]{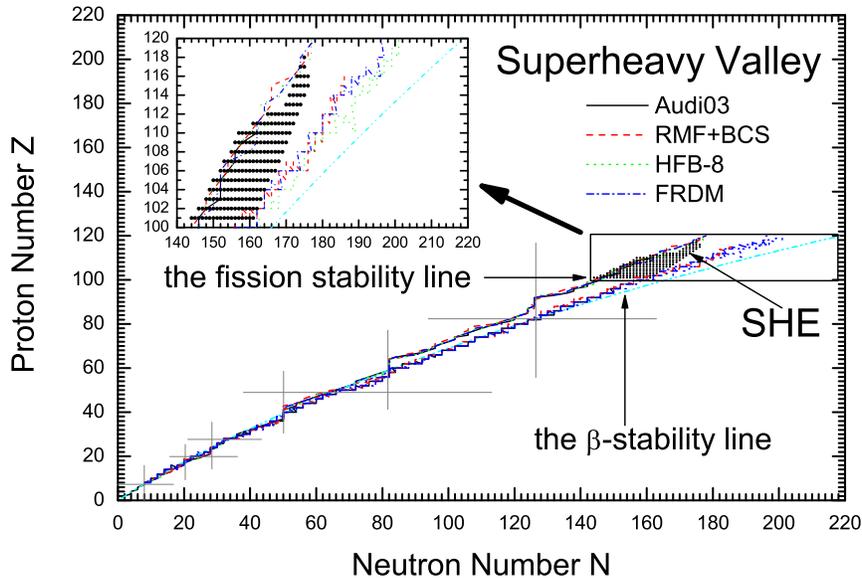}
\caption{(color online). The fission stability line and the
$\beta$-stability line as functions of neutron number $N$ and
proton number $Z$. The theoretical predictions (RMF+BCS, HFB-8
\cite{Samyn04} and FRDM \cite{Moller95}) are compared with
existing experimental data (including those obtained using
``systematic trends") \cite{Audi03}. The smooth dash-dot-dot line
is the $\beta$-stability line of Bohr-Mottelson \cite{Bohr98} (see
text). The small black dots represent experimentally synthesized
superheavy nuclei (SHE) (including those obtained using
``systematic trends") \cite{Audi03}. \label{sline}}
\end{figure*}

It was argued from the energy point of view that the fission
stability line, the line connecting the nucleus with maximum
binding energy per nucleon in each isotopic chain, plays an
important role in studies of the stability of the heaviest nucleus
in Ref. \cite{Wu96}. In Fig. \ref{mass} the binding energies per
nucleon of the 600 nuclei we calculated are plotted as functions
of the neutron number $N$ (upper panel). The experimental data
(lower panel) are taken from Ref. \cite{Audi03}. The nucleus with
maximum binding energy per nucleon in each isotopic chain is
denoted by a triangle. Those nuclei in different isotopic chains
are then connected by dashed lines. It can be clearly seen that
the area we investigated has in fact included the most bound
nucleus in each isotopic chain. The theoretical curves are almost
identical to the experimental ones (see also Fig. \ref{sline}).
Therefore, just as expected, all the experimental syntheses are
indeed near the fission stability line \cite{Wu96}. Since the
results of the HFB-8 \cite{Samyn04} and FRDM \cite{Moller95} mass
formulae are very similar to our calculations, they are not shown
in this figure. Here, a few words about HFB-8 and FRDM are in
place. The mass table of the finite-range droplet model has been
around for more than ten years \cite{Moller95}. By carefully
adjusting its parameters (about thirty) to the saturation
properties of nuclear matter and the binding energies of around
1000 nuclei, it obtained a root-mean-square deviation of about 0.6
MeV for the binding energies of all the experimentally-known
nuclei. HFB-8 is a rather new mass table based on the
Hartree-Fock-Bogoliubov method \cite{Samyn04}. By performing
particle number projection, incorporating phenomenologically both
the Wigner energy and the rotational energy, and meanwhile
adjusting its parameters (around 20) to fit the saturation
properties of nuclear matter and the binding energies of about
2000 nuclei, it achieved a similar quality to that of the FRDM
model. More comparisons between these two models and current RMF
models for nuclear ground-state properties can be found in Ref.
\cite{Geng05}.

In Fig. \ref{sline}, the fission stability line and the
$\beta$-stability line are plotted as functions of $N$ and $Z$.
The three theoretical predictions (RMF+BCS, HFB-8 \cite{Samyn04}
and FRDM \cite{Moller95}) are compared with existing experimental
data \cite{Audi03}. For reference, the phenomenological
$\beta$-stability line
 \begin{equation}\label{fsl}
 N-Z=6.0\times 10^{-3}A^{5/3}
 \end{equation}
 derived from the Bethe-Weizs\"{a}cker mass formula in Ref. \cite{Bohr98} is also shown.
 Several interesting things can be learned immediately from Fig.
\ref{sline}. First, the microscopic (RMF+BCS, HFB-8 and FRDM)
$\beta$-stability lines agree with
 each other very well, but they are bent up a little bit compared to the phenomenological $\beta$-stability line
 of Bohr-Mottelson. This difference can slightly change the definition of proton
richness or neutron richness. As we can see from
 Fig. \ref{sline}, superheavy nuclei have reached the $\beta$-stability line around $Z=100$ if we use the microscopic
 $\beta$-stability lines. However, there is still a gap if we use the phenomenological $\beta$-stability line.
 In the following discussions, we will not distinguish the microscopic
$\beta$-stability
 line and the phenomenological one. Second, all the three
 modern models predict quite similar most bound nuclei in each isotopic chain, which
 agree remarkably well with the experimental data. Third, the
 experimentally-synthesized superheavy nuclei populate to the left of the fission stability line and
 to the right of the $\beta$-stability line, the same as what we observed for those heavy nuclei with
 $89\le Z\le 100$ \cite{Geng05}. This finding leads us to speculate that
 the center of the superheavy island is most probably in between these two lines. We will call this region the ``superheavy valley".  This
speculation can find its support in ``ordinary nuclei'', where we
know
 most nuclei exist along the $\beta$-stability line. But as we can see from Fig. \ref{sline}, for them
 the fission stability line is quite close to the $\beta$-stability line. For heavier nuclei, such as
 actinide nuclei, they populate in between \cite{Geng05}. However,
 since we have not taken into account alpha-decay
  in this work, the exact location where the superheavy
 nucleus has the longest lifetime cannot be predicted.
 Lastly, the shell effects lead the most bound nucleus in each isotopic
chain in the superheavy region to tend to have more protons,
compared to their light- or medium-mass counterparts. The special
role of neutron shell closure in this respect is most conspicuous
at $N=50$, 82 and 126. After
 three major neutron shell closures, the fission stability line
 deviates a lot from the $\beta$-stability line in the superheavy
 region, and thus, forms the so-called ``superheavy valley''.

 The last observation above endows us with a powerful tool of identifying the next
 major neutron shell closure. That is to say there should also be a
 sudden increase in the fission stability line at the next major
 neutron shell closure. Using this argument, no neutron shell closure is found in
 the area we investigated, i.e. $N\le180$. It would be very interesting to
 study the $N>180$ region, but it might be a difficult task due to
 the amount of computer resources needed. Fortunately, using our
 findings here, the searching region can be considerably reduced.
 Noticing that the fission stability line behaves as a linear function of $N$ and $Z$ between two
 major neutron shell closures, we can approximate the theoretical and
 experimental fission stability lines (for $N>126$) by the following formulae:
 \begin{eqnarray}
 Z&=&(17.105\pm0.959)+(0.577\pm0.006)N\;\mbox{(theory)},\\
 Z&=&16.118+0.580 N\;\mbox{(Audi03)}.
 \end{eqnarray}
The fission stability lines derived from the RMF+BCS, HFB-8 and
FRDM calculations are all quite similar to each other, and
therefore they have been averaged to obtain the above formula
denoted by ``theory''. One should note that these approximations
are valid only up to the next major neutron shell closure. Future
investigations, therefore, can be performed along these lines with
the fission stability line as the upper bound.

\section{Shapes of superheavy nuclei}

\begin{figure}[t]
\includegraphics[scale=0.4]{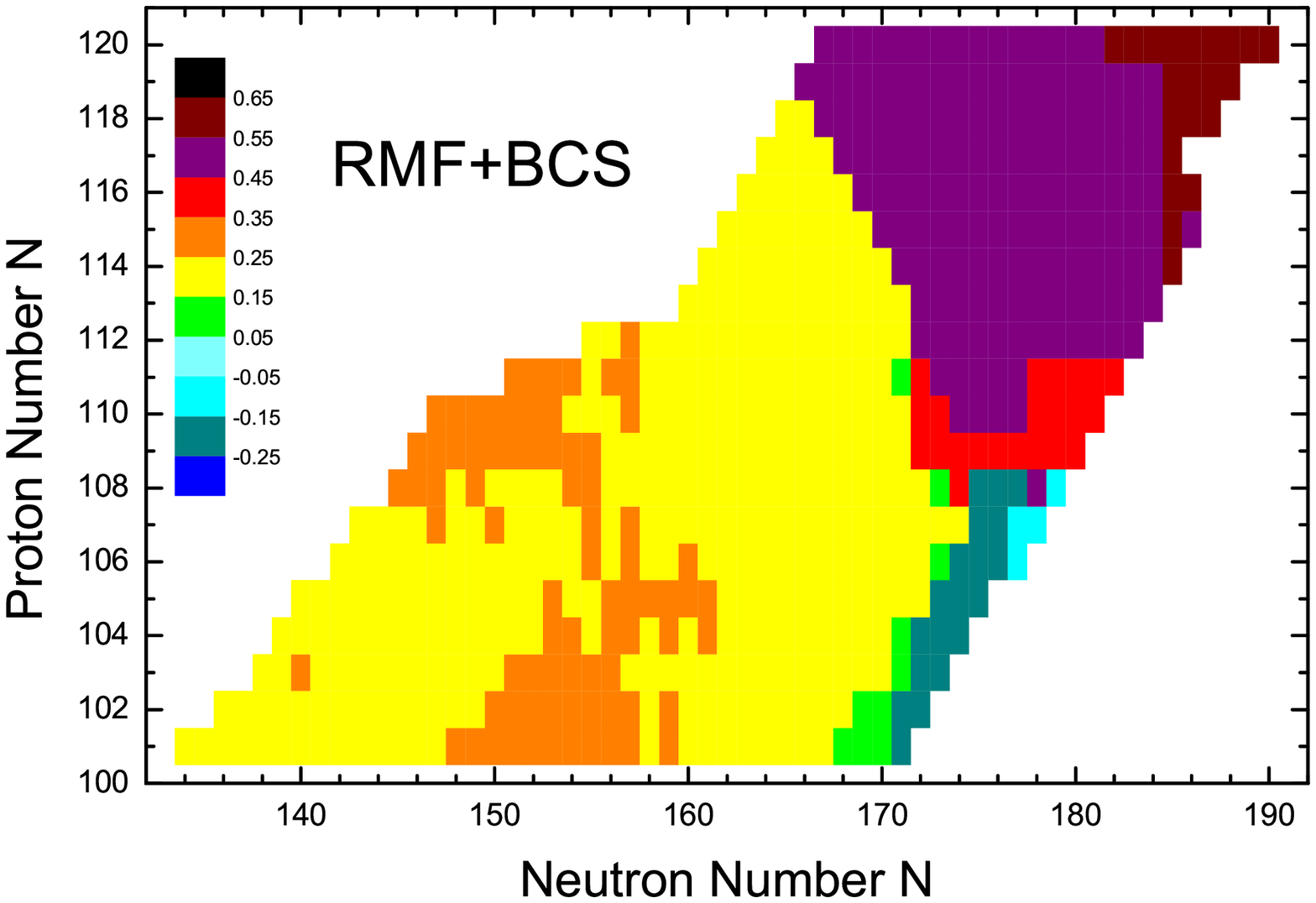}
\includegraphics[scale=0.4]{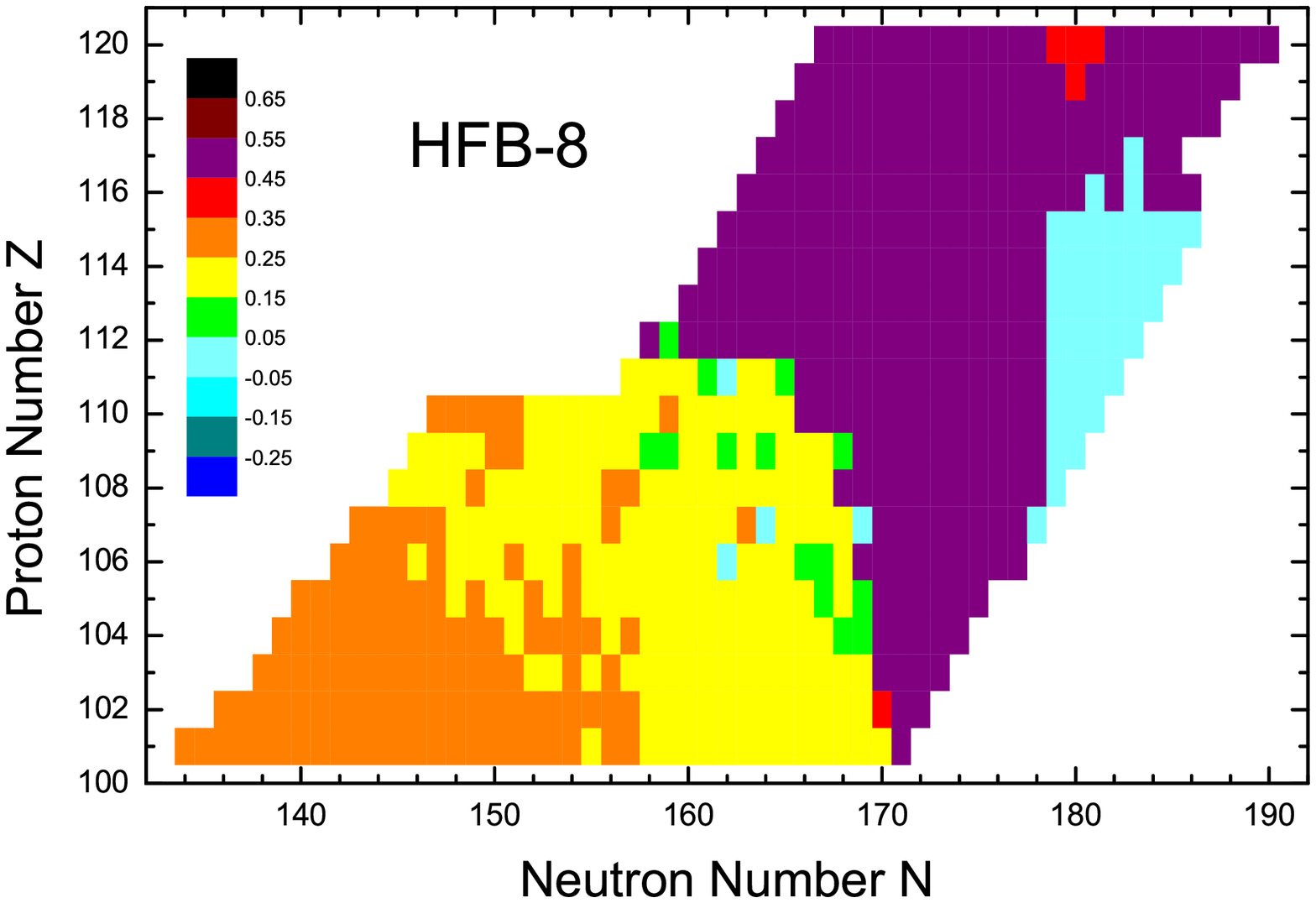}
\includegraphics[scale=0.4]{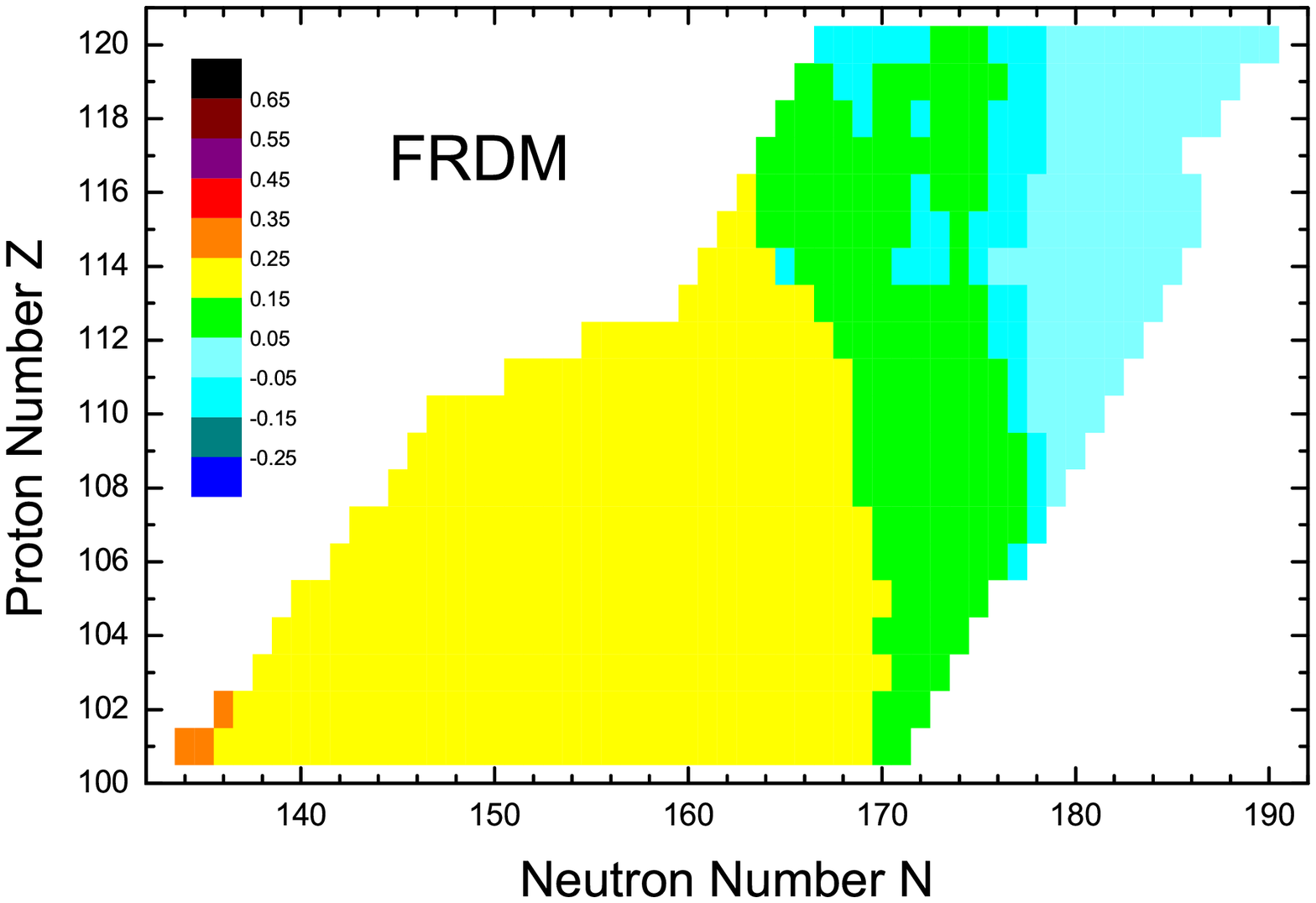}
\caption{(color online). Proton quadrupole deformation parameters,
$\beta_2$, of superheavy nuclei with $Z = 101$-120 and $N =
134$-190 as functions of neutron number $N$ and proton number $Z$. The predictions of the
RMF+BCS calculations are compared with those of the HFB-8
\cite{Samyn04} and FRDM \cite{Moller95} mass
formulae.\label{beta}}
\end{figure}
Most experimentally-synthesized superheavy nuclei are believed to
be deformed. This could be verified from two aspects: First,
studies of ``ordinary nuclei" revealed that only magic nuclei are
spherical; second, actinide nuclei are known to be strongly
deformed. In Fig. \ref{beta}, we plot the quadrupole deformation
parameters, $\beta_2$, of the 600 nuclei we calculated as
functions of neutron number $N$ and proton number $Z$. It is quite
interesting to note that both RMF+BCS and HFB-8 calculations
predict very strongly prolate deformations, $\beta_2\ge0.45$, for
those nuclei in the upper-right corner. Even for those nuclei in
the lower-left corner, the deformation is still appreciable, with
$\beta_2$ ranging from 0.15 to 0.35. However, there are two major
differences between the predictions of the RMF+BCS calculations
and those of the HFB-8 calculations. Firstly, there is a small
number of spherical nuclei near $N = 184$ and $Z = 114$ in the
HFB-8 calculations, but there is no spherical nucleus at all in
the RMF+BCS calculations. Second, strongly prolate deformation
occurs in the HFB-8 calculations at mass numbers smaller than
those predicted by the RMF+BCS calculations; therefore it might
indicate possible shape coexistence
\cite{Geng03PRC,Ren03,Cwiok05}.

In other words, the RMF model with the effective force TMA
predicts no doubly magic nucleus in the area we investigated. The
HFB-8 calculations seem to prefer $^{298}_{184}$114 to be a doubly
magic nucleus, but the deformation pattern predicted for these
superheavy nuclei are different from those observed for ``ordinary
nuclei" \cite{Geng05,Geng05PTP}, where a much larger number of
nuclei near the doubly magic nucleus are found to be spherical.
Therefore a decisive conclusion is not possible. This is, in fact,
consistent with the recent study of \'{C}wiok et al.
\cite{Cwiok05}, where they discussed the possibility of triaxial
deformations for those nuclei near $N = 184$ and $Z = 114$. Thus,
this difference is once again the old model dependence problem. In
Ref. \cite{Zhang04}, the spherical doubly magic nuclei are
searched for all the usual parameterizations while here TMA is
used to investigate their deformations. One should note that the
octupole deformation and other beyond mean-field correlations can
greatly reduce the second fission barrier \cite{Burvenich04PRC};
therefore, it would be interesting to investigate whether the
conclusions of the present work would be modified if these
correlations are taken into account.

It is to be noted that the most remarkable difference exists
between the microscopic models, RMF+BCS and HFB-8, and the
macroscopic-microscopic model, FRDM: FRDM displays a transition
from moderately deformed shapes to spherical shapes from the
lower-left corner to the upper-right corner. That is to say we
find a similar pattern that we observed for those conventional
doubly magic nuclei \cite{Geng05,Geng05PTP}, i.e. spherical shapes
for doubly magic nuclei and those nearby. Therefore once again we
conclude that $^{298}_{184}114$ is probably a doubly magic nucleus
in the FRDM calculations, or more precisely, there is probably a major shell
closure at $N=184$.

Of course, one should note that no clear indication of the $N=184$
shell closure in the RMF+BCS and HFB-8 calculations (as shown in
Fig.\ref{beta}) does not necessarily mean it does not exist. As
well known nowadays, proton (neutron) shell closures are also
neutron (proton) number dependent. Therefore, this implies if
$N=184$ is truly a neutron shell closure,  we might have to look
for the evidence in the region with $Z>120$. Similarly,  $Z=114$
or $Z=120$ might be a major proton shell closure in the more
neutron rich side.

\section{Summary}

By performing a RMF+BCS calculation of about 600 superheavy nuclei
and employing the latest theoretical and experimental results, we
have studied two extremely important subjects of superheavy
nuclei, their stabilities and shapes. Both theory and experiment
showed that all the experimentally-synthesized superheavy nuclei
lie in between the fission stability line and the
$\beta$-stability line, i.e. the ``superheavy valley". It was also
shown that the fission stability line and the $\beta$-stability
line tend to be more proton rich than their light- or medium-mass
counterparts. In this sense, it is justified to say that the
observed ``proton richness'' of superheavy nuclei is not only a
result of the limitation of current experimental methods but also
a manifestation of their inherent nature. Although all the three
theoretical models (RMF+BCS, HFB-8 and FRDM) predict most
superheavy nuclei to be deformed, they differ from each other for
those nuclei near $N=184$. This model dependence might be removed
by a more precise description of the single-particle spectra of
the heaviest nuclei, which would be our next work.

L. S. Geng acknowledges the fruitful discussions with Dr. J. Meng,
Dr. S.-G. Zhou and Dr. H. F. L\"{u}.
%

\newpage

\end{document}